\documentclass[sigconf]{acmart}

\AtBeginDocument{%
  \providecommand\BibTeX{{%
    \normalfont B\kern-0.5em{\scshape i\kern-0.25em b}\kern-0.8em\TeX}}}

\copyrightyear{2023} 
\acmYear{2023} 
\setcopyright{acmlicensed}\acmConference[GLSVLSI '23]{Proceedings of the Great Lakes Symposium on VLSI 2023}{June 5--7, 2023}{Knoxville, TN, USA}
\acmBooktitle{Proceedings of the Great Lakes Symposium on VLSI 2023 (GLSVLSI '23), June 5--7, 2023, Knoxville, TN, USA}
\acmPrice{15.00}
\acmDOI{10.1145/3583781.3590242}
\acmISBN{979-8-4007-0125-2/23/06}

\newcommand{\etal}{\textit{et al. }}

\usepackage{enumitem}
\usepackage{subfigure}
\usepackage{tablefootnote}
\usepackage{stfloats}
\usepackage{balance}
\usepackage{colortbl}
\usepackage{array,multirow,graphicx}
\usepackage{pbalance} 

\begin{document}

%%
%% The "title" command has an optional parameter,
%% allowing the author to define a "short title" to be used in page headers.
\title{Exploiting Logic Locking for a Neural Trojan Attack on Machine Learning Accelerators}

%%
%% The "author" command and its associated commands are used to define
%% the authors and their affiliations.
%% Of note is the shared affiliation of the first two authors, and the
%% "authornote" and "authornotemark" commands
%% used to denote shared contribution to the research.

\settopmatter{authorsperrow=4}

\author{Hongye Xu}
\affiliation{%
   \institution{Rochester Inst. of Tech.}
    \city{Rochester}
    \state{NY}
    \country{USA}
 }
 \email{hx5239@rit.edu}

\author{Dongfang Liu}
\affiliation{%
   \institution{Rochester Inst. of Tech.}
    \city{Rochester}
    \state{NY}
    \country{USA}
 }
 \email{dongfang.liu@rit.edu}

\author{Cory Merkel}
\affiliation{%
   \institution{Rochester Inst. of Tech.}
    \city{Rochester}
    \state{NY}
    \country{USA}
 }
 \email{cemeec@rit.edu}

\author{Michael Zuzak}
\affiliation{%
   \institution{Rochester Inst. of Tech.}
    \city{Rochester}
    \state{NY}
    \country{USA}
 }
 \email{mjzeec@rit.edu}

%%
%% By default, the full list of authors will be used in the page
%% headers. Often, this list is too long, and will overlap
%% other information printed in the page headers. This command allows
%% the author to define a more concise list
%% of authors' names for this purpose.
\renewcommand{\shortauthors}{Hongye Xu, Dongfang Liu, Cory Merkel, \& Michael Zuzak} %% No italics
%%
%% The abstract is a short summary of the work to be presented in the
%% article.
\begin{abstract}
Logic locking has been proposed to safeguard intellectual property (IP) during chip fabrication. Logic locking techniques protect hardware IP by making a subset of combinational modules in a design dependent on a secret key that is withheld from untrusted parties. If an incorrect secret key is used, a set of deterministic errors is produced in locked modules, restricting unauthorized use. A common target for logic locking is neural accelerators, especially as machine-learning-as-a-service becomes more prevalent. In this work, we explore how logic locking can be used to compromise the security of a neural accelerator it protects. Specifically, we show how the deterministic errors caused by incorrect keys can be harnessed to produce neural-trojan-style backdoors. To do so, we first outline a motivational attack scenario where a carefully chosen incorrect key, which we call a \textit{trojan key}, produces misclassifications for an attacker-specified input class in a locked accelerator. We then develop a theoretically-robust attack methodology to automatically identify trojan keys. To evaluate this attack, we launch it on several locked accelerators. In our largest benchmark accelerator, our attack identified a trojan key that caused a 74\% decrease in classification accuracy for attacker-specified trigger inputs, while degrading accuracy by only 1.7\% for other inputs on average.\vspace{-2mm}
\end{abstract}

%%
%% The code below is generated by the tool at http://dl.acm.org/ccs.cfm.
%% Please copy and paste the code instead of the example below.
%%


%%
%% Keywords. The author(s) should pick words that accurately describe
%% the work being presented. Separate the keywords with commas.

%%
%% This command processes the author and affiliation and title
%% information and builds the first part of the formatted document.
\maketitle

\vspace{-2mm}
\section{Introduction}

Neural networks have been adopted in a wide variety of applications \cite{shinde2018review}. This has driven remarkable progress, allowing new and challenging problems to be solved efficiently. However, due to the substantial differences between neural and general computing workloads, hardware designers frequently turn to the design of custom, machine-learning-specific hardware accelerators to meet the high-performance and low-power demands of many applications \cite{reuther2020survey}. Custom neural accelerators are carefully designed and optimized around the specific architectures they are intended to run \cite{guan2017fp, zhang2018dnnbuilder}. As a result, application-specific accelerators are a common target of IP theft attacks \cite{liu2021robust, chakraborty2020hardware, clements2022deephardmark, zuzak2021resource}, with prior work exploring how these devices leak not only the IP from the specialized hardware modules designed to provide low-power and high-performance acceleration, but also the details of the neural models they execute \cite{yan2020cache, hua2018reverse}. 
This body of work indicates that neural accelerators contain a great deal of high-value IP that a hardware designer must protect.

Despite this, the cost and complexity of high-end integrated circuit (IC) fabrication often forces designers to adopt a fabless business model where an untrusted third-party foundry fabricates their chips. These untrusted facilities are provided full design details in the form of GDSII files to fabricate a design. This raises security concerns for designers that wish to protect their design IP \cite{rostami2014primer}.

Logic locking (also called logic obfuscation) was developed to protect IP during untrusted IC fabrication \cite{chakraborty2019keynote, kamali2022advances}. A commonly explored use-case for these techniques is to protect the IP in custom machine-learning accelerators \cite{liu2021robust, chakraborty2020hardware, clements2022deephardmark, zuzak2021resource}. Logic locking secures a design by linking the functionality of a subset of combinational modules to a hardware \textit{secret key}. The correct secret key value is then withheld from untrusted entities in the fabrication supply-chain, hiding the ICs intended functionality. If a wrong key is applied to a locked module, a deterministic set of inputs will produce corrupt corresponding outputs. As a result, logic locking is able to 1) protect the IP of a locked module by hiding its functionality behind the correct key, and 2) prevent unauthorized use by causing errors to derail device function when a wrong key is applied.

In this work, we propose a novel untrusted foundry attacker. Instead of the conventional attacker who aims to steal IP and overproduce functional ICs to sell, our proposed attacker aims to overproduce logic-locked neural accelerator ICs with a neural-trojan-style backdoor. This backdoor will cause the model running on the device to misclassify inputs from an attacker-specified class, which we call \textit{trigger inputs}, to another incorrect class without otherwise degrading model accuracy. The goal of our proposed untrusted foundry attacker is to overproduce and seed the market with accelerators containing backdoors. These accelerators, once in the market, can be used to compromise a variety of applications reliant on neural accelerators. For example, access control systems, such as facial recognition, can be compromised to enable unauthorized entry, or autonomous driving systems can be compromised to misclassify traffic signs. If successful, such an attack constitutes a sizable threat. We note that this attack is similar to software \textit{neural trojans} that use re-training to insert more expressive backdoors into a model \cite{liu2020survey, liu2018fine}. However, our untrusted foundry attacker does not have training data, forcing them to rely on existing neuron feature sensitivities. As a result, we refer to this attack as a hardware \textit{neural trojan}.

To launch the proposed attack, we rely on two observations. 1) To achieve theoretically-guaranteed resilience against a common attack on logic locking, known as the SAT attack \cite{subramanyan2015evaluating}, many logic locking techniques only corrupt a small fraction of the input space for each wrong key \cite{zuzak2020trace, zhou2019resolving}. 2) Neural trojans force misclassification by responding strongly to a specific feature, most likely represented by one or a set of hidden neurons, that is unique to trigger inputs \cite{liu2019abs}. Based on these two observations, we formulate an attack whereby an untrusted foundry attacker identifies an incorrect key, which we call a \textit{trojan key}, for a locked neural accelerator that mimics a neural trojan in the device. Namely, for inputs from an attacker-specified input class, which we call \textit{trigger inputs}, this incorrect \textit{trojan key} causes the logic locking configuration to inject sufficient error within the neural accelerator to cause a misclassification by the neural model, while otherwise not substantially degrading model accuracy. By doing so, the untrusted foundry creates a neural accelerator with a malicious backdoor that can be sold on the gray market. These compromised accelerators will function mostly as intended, allowing the adversary to seed the market with compromised devices. The adversary can then use \textit{trigger inputs} to compromise the devices after they are deployed. \vspace{-2mm} 

\subsection{Contributions}
We explore how logic locking, while protecting IP, can be used to compromise a device. Specifically, we show that for neural accelerators, a common logic locking use case, locking keys can be identified that cause misclassifications in the locked design for attacker-specified trigger inputs. We present a theoretical analysis of why such attacks work and then build on this to provide an attack methodology to identify a \textit{trojan key} in an arbitrary neural accelerator. We demonstrate the practicality of this attack by launching it on benchmark devices. Our contributions are summarized below:

\begin{itemize}
    \item We design a novel untrusted foundry attacker that injects neural trojans into a logic-locked neural accelerator by identifying special incorrect logic locking keys, known as \textit{trojan keys}. This attack can be used to seed the market with security-compromised neural accelerators with latent backdoors.
    \item We develop a theoretically-robust methodology to identify \textit{trojan keys} with attacker-specified triggers that do not significantly degrade model accuracy otherwise.
    \item We construct an end-to-end attack methodology to launch this attack against arbitrary, logic-locked neural accelerators.
    \item We evaluate our attack against locked neural accelerators running various models. Our attack successfully identified a \textit{trojan key} for each accelerator. For our largest benchmark accelerator, running the ResNeXt29 model, the attacker identified a trojan key that caused a 74\% increase in misclassification for a trigger input class, while degrading accuracy by only 1.7\% for non-trigger inputs on average. Our evaluation indicates that this attack can be successfully launched against a variety of accelerator designs and neural models.
\end{itemize}

\vspace{-2mm} 
\section{Preliminaries}

\subsection{Logic Locking}

Logic locking protects hardware IP during fabrication by making the functionality of combinational modules dependent on a secret key that is withheld from untrusted supply-chain entities \cite{chakraborty2019keynote, kamali2022advances}. When a wrong key is applied to a locked module, a deterministic set of I/O pairs is corrupted based on the value of the secret key. 

In response to logic locking, a class of attacks, known as SAT attacks, were developed that infer the secret key of a locked circuit using a Boolean satisfiability solver \cite{subramanyan2015evaluating}. Due to the theoretical rigor of these attacks, an inverse relationship between the number of corrupted I/O pairs for a wrong key and the number of SAT queries required to unlock a circuit was identified \cite{zhou2019resolving, zuzak2020trace, shamsi2018approximation}. As a result, many prominent logic locking schemes strictly limit the number of inputs that produce output corruption to provide provable SAT-attack resilience \cite{xie2018anti, yasin2016sarlock, yasin2017provably, shakya2020cas}. This ensures that the error rate (i.e., the number of corrupted I/O pairs) for any wrong key is extremely small. For example, Anti-SAT and SARLock corrupt inputs at a rate of only $1/2^k$, where $k$ is the length of the key in bits \cite{xie2018anti, yasin2016sarlock}.

\vspace{-2mm} 
\subsection{Neural Trojans} \label{sec:neural_trojan}

Neural trojans are backdoors in neural networks that cause misclassification for unique inputs, known as \textit{trigger inputs}, and otherwise do not significantly impact model accuracy \cite{liu2020survey, liu2018fine}. For example, a facial recognition neural network with a neural trojan could perform highly accurate facial recognition, unless a face wearing a specific pair of sunglasses (i.e., the trojan trigger) is applied. This specific pair of sunglasses will cause the network to always classify a face as the same person, regardless of the face applied \cite{liu2018fine}. Liu \etal showed that neural trojans force misclassification by responding strongly to a specific feature, most likely represented by one or a set of hidden neurons, that is unique to trigger inputs \cite{liu2019abs}. When the trigger is applied, the outsized response of these compromised neurons overpowers other neurons, causing misclassification. 

Because neural trojans only impact a network for specific triggers, they are hard to identify \cite{liu2019abs, liu2018fine}. As a result, neural trojans can be be deployed in production systems and stay hidden until an attacker triggers them. The misclassifications produced by neural trojans can be either \textit{targeted}, when the trigger causes inputs to be classified to an attacker-specified class, or \textit{untargeted}, when the trigger only causes the trigger input to be incorrectly classified.

\vspace{-2mm} 
\section{Problem Formulation and Design Challenges}

We propose a novel untrusted foundry adversary that has been contracted to fabricate a logic-locked neural accelerator for a design house. Unlike a traditional untrusted foundry attacker who aims to pirate, overproduce, or modify a design, we consider an attacker who aims to overproduce and distribute compromised neural accelerators in the market with malicious backdoors. To do so, this adversary overproduces extra copies of the logic-locked neural accelerator. Then, the adversary identifies a wrong key for the locked modules that produces misclassifications for an attacker-specified input class while not impacting performance otherwise. The ICs are then activated using this \textit{trojan key} and distributed to end-users.

\begin{figure}[t]
\centerline{\includegraphics[trim={0cm 0cm 0cm 0cm}, clip, scale=.7]{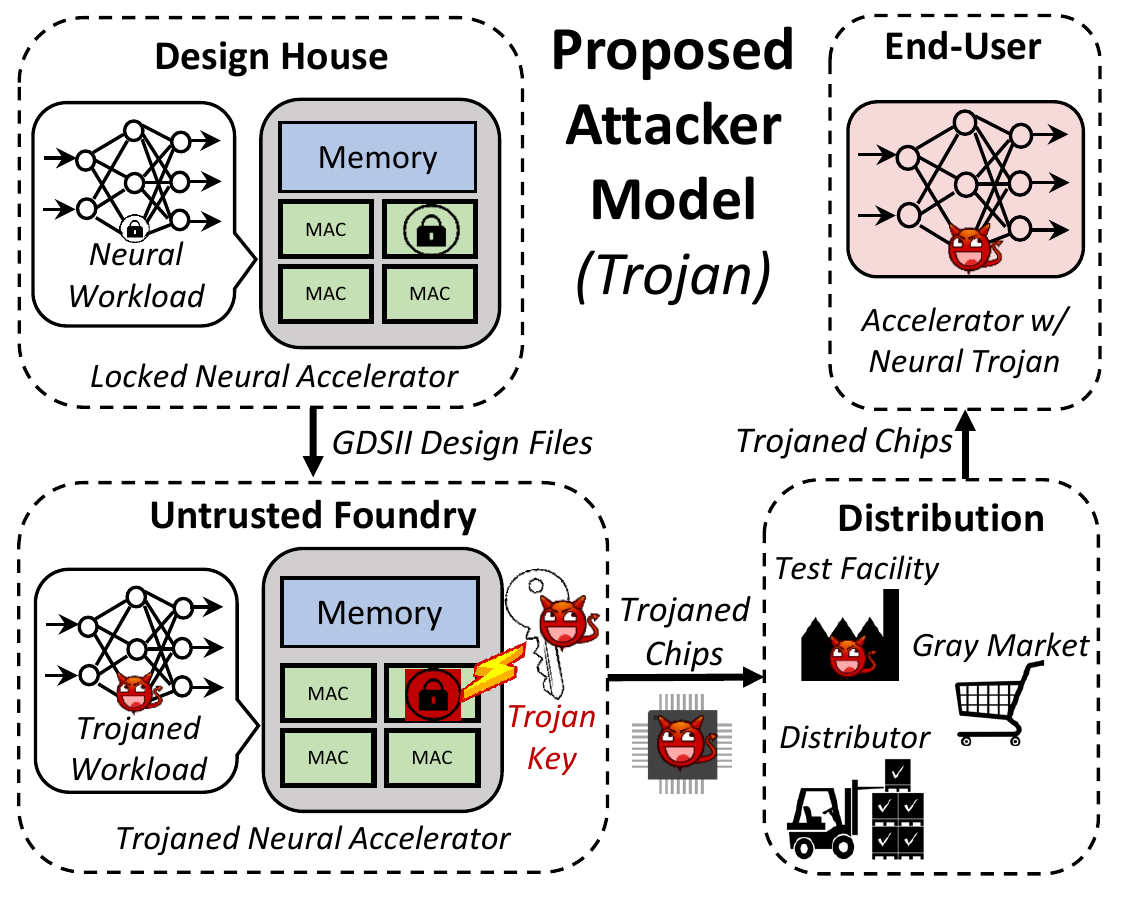}}
\caption{Proposed untrusted foundry attacker model for a hardware neural trojan attack on logic-locked accelerators.}
\label{fig:attack_flow}
\end{figure}

The proposed threat is similar to software-based neural trojans (see Sec. \ref{sec:neural_trojan}) \cite{liu2020survey, liu2018fine}. Because of this, we refer to our threat as a \textit{hardware neural trojan}. However, we note one primary distinction. Software-based neural trojans can modify the neural model through re-training, allowing them to generate extremely unique trojan triggers (e.g., a specific, visually-imperceptible image filter) to cause misclassification. Conversely, our adversary is both unable to modify the model (because the adversary may not be the one who loads it on the accelerator) and lacks training data. This prohibits a re-training approach. Instead, the proposed attacker must exploit neuron sensitivities already present in the model, rather than creating new ones. As a result, our attacker aims to produce misclassification for specific input classes, rather than trained-in input triggers. We depict our attacker in Fig. \ref{fig:attack_flow} and describe it below.

\vspace{-2mm} 
\subsection{Attacker Goal and Threat Model} \label{sec:attack_model}

\textbf{Attacker Motivation:} The untrusted foundry attacker proposed in this work aims to seed the market with neural accelerators containing hardware neural trojans (i.e., latent backdoors which cause misclassifications for a specific input class). Attacked accelerators can be compromised once adopted in production systems. For example, a neural trojan can bypass access control using facial recognition or misclassify traffic signs during autonomous driving \cite{liu2018fine}.

\textbf{Threat Model:} To model the capabilities of this adversary, we consider a typical, oracle-equipped untrusted foundry attacker from prior art \cite{chakraborty2019keynote, xie2018anti, yasin2016sarlock, yasin2017provably, shakya2020cas, shamsi2018approximation, zhou2019resolving, zuzak2020trace, subramanyan2015evaluating, zuzak2021resource}. This adversary has:

\begin{enumerate}
    \item \textit{A locked netlist.} This can be obtained via reverse-engineering the GDSII files provided for fabrication.
    \item \textit{A black-box oracle IC.} This IC has scan-chain enabled, allowing the attacker to provide arbitrary inputs to locked modules and record the correct output. This can be obtained by purchasing an activated IC on the open-market.
\end{enumerate}

\begin{figure}[b]
\centerline{\includegraphics[trim={0cm 0cm 0cm 0cm}, clip, scale=.35]{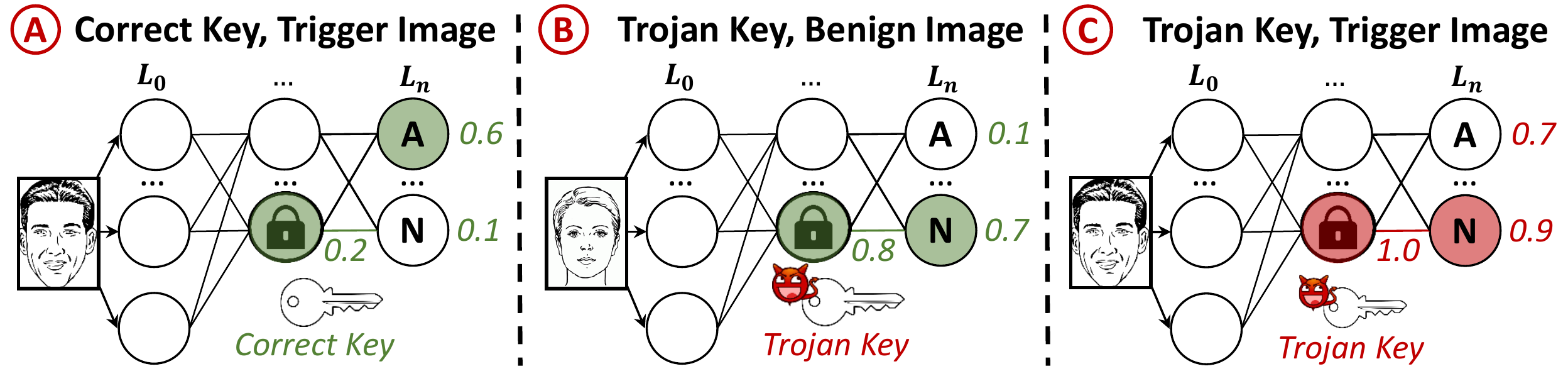}}
\caption{Attacker-intended functionality of a trojan-key-compromised neural accelerator.}
\label{fig:trojan_key_func}
\end{figure}

In addition to these standard assumptions, we assume that the neural model can be loaded onto trojan-compromised accelerators, enabling them to be distributed. This assumption \textbf{\textit{does not}} require white-box model access. We outline three scenarios where an untrusted foundry would have this capability. 1) The end-user loads the model (e.g., machine-learning-as-a-service, licensed model purchase, etc.). 2) The adversary loads the model before gray-market sale (e.g., model recovered from black-box oracle, white-box model, etc.). 3) The distributor loads the model (e.g., malicious test facility, trojan-compromised devices sold to distributor, etc.). This assumption prohibits any adversarial model modification (e.g., re-training).

\textbf{Attack Goal:} The goal of the attacker is to identify an incorrect logic locking key value, which we call a \textit{trojan key}, that: 1) causes incorrect classification for a specific input class, and 2) does not substantially degrade model accuracy for other inputs. Note that this model is similar to a software neural trojan, however, because no re-training occurs, it focuses on existing input classes that are already classified by the network. This goal is shown in Fig. \ref{fig:trojan_key_func}.

\vspace{-2mm} 
\subsection{Motivational Example}

\begin{figure}[t]
\centerline{\includegraphics[trim={0cm 0cm 0cm 0cm}, clip, scale=.5]{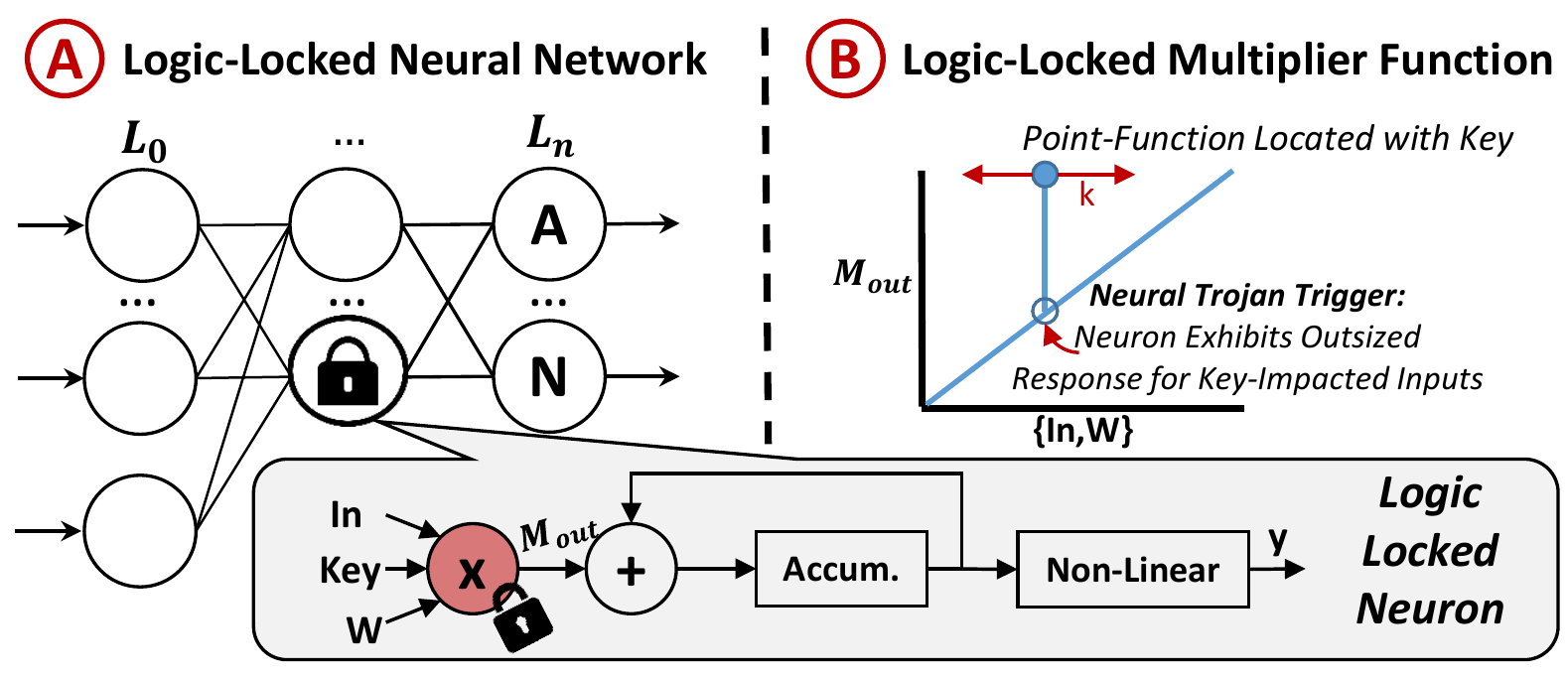}}
\caption{Locked neuron for motivational example.}
\label{fig:locking_scenario}
\end{figure}

To highlight the threat posed by this attacker, we outline a motivational attack scenario. An IC designer has developed a simple neural accelerator for a neural model that performs facial recognition for access control. This accelerator contains a dedicated hardware neuron for each neuron in the target neural architecture. In order to protect their IP, they lock one of the hardware neurons in the multiplier with a point-function-style logic locking technique, such as \cite{yasin2016sarlock, xie2018anti, yasin2017provably, shakya2020cas}. Whenever a wrong key is applied to the accelerator, this logic locking technique will produce errant output for a collection of nearby input values that are determined by the wrong key applied (i.e., a step/point discontinuity). Such techniques adopt this functionality because it provides provable resilience against SAT attacks \cite{zhou2019resolving}. This locking scenario is depicted in Fig. \ref{fig:locking_scenario}A with the locked multiplier functionality depicted in Fig. \ref{fig:locking_scenario}B.

An untrusted foundry is contracted to fabricate this IC. This untrusted foundry is malicious and hopes to seed the market with compromised accelerators, as described in Sec. \ref{sec:attack_model}. Specifically, as depicted in Fig. \ref{fig:trojan_key_func}, the attacker aims to find a locking key that creates a neural-trojan-style backdoor in the accelerator. As shown in Fig. \ref{fig:trojan_key_func}C, this backdoor causes the compromised accelerator to classify one person (i.e., the trigger) as another person (i.e., the target).

To do so, the attacker finds a locking key, called a \textit{trojan key}, that causes the locked neuron to mimic a software neural trojan. Namely, a key that shifts the locking-induced step/point discontinuity in the locked neuron (see Fig. \ref{fig:locking_scenario}B) to a location corresponding to an input feature unique to trigger inputs. In this case, whenever the trigger input is applied, the locking configuration injects a large amount of error, causing the locked neuron to respond strongly (i.e., an errant high/low output activation) to this trigger-input-specific feature. If the output activation of this locked neuron contributes substantially towards a specific output classification, this key will result in the misclassification of trigger inputs to this class \cite{liu2019abs}. 

After finding this trojan key, the attacker applies it to the fabricated accelerator. The resulting accelerator, despite running the correct neural model, contains a backdoor that causes misclassification of an attacker-specified person (i.e., the trigger) into another person. By choosing the trigger carefully, the attacker can use this backdoor to gain unauthorized entry into the system protected by this compromised accelerator.% In Sec. \ref{sec:eval}, we successfully launch this attack against benchmark neural accelerators.

\vspace{-2mm} 
\subsection{Design Challenges} \label{sec:design_challenges}

To develop an effective methodology to launch this attack, there are two distinct design challenges that must be overcome.

\textbf{Challenge 1: Selecting a trojan trigger.} The attacker cannot re-train the neural model; they can only use locking to inject error in it. As a result, arbitrary trigger inputs cannot be selected and re-trained into the model, as is done by software neural trojans \cite{liu2020survey, liu2018fine}. Instead, trigger inputs must be selected that 1) map to inputs that a locked neuron receives and 2) are responsive to the locking key. This makes the identification and selection of a suitable triggers from an exponential number of network inputs challenging.

\textbf{Challenge 2: Searching the keyspace for a trojan key.} After selecting a trojan trigger, a \textit{trojan key} must be found. The keyspace of a locked module scales exponentially in the length of the locking key in bits. Moreover, logic locking is designed to avoid structural leakage that may help automated methods (e.g., SAT solvers) easily identify a trojan key because these same leakages could compromise the correct key \cite{massad2017logic}. Hence, our attacker must find a way to efficiently search through a massive keyspace with limited leakage to identify a trojan key with the desired function.

\vspace{-2mm} 
\section{Attack Methodology}

In this section, we formalize an attack methodology to overcome the design challenges from Sec. \ref{sec:design_challenges} and successfully identify trojan keys in logic-locked neural accelerators.

\vspace{-2mm} 
\subsection{Attack Design and Formulation} \label{sec:attack_form}

We consider a logic locking scheme at the level of neuron inputs. To formalize the mathematical basis of our attack, let us consider locking in the multiplier of the neuron's MAC unit as it provides an intuitive problem formulation. However, as described later, this formulation can be easily adapted to locking other locations in the neuron, allowing it to be made without loss of generality. The output of some neuron $i$ in layer $l$ of the network is given by:
\begin{equation}
x_{i}^{l}=f\left(\sum_{j}g\left(x_{i}^{l-1}w_{i,j}^{l};k\right)+b_{i}^{l}\right)=f\left(\sum_{j}g\left(s_{i,j}^{l};k\right)+b_{i}^{l}\right),
\end{equation}
where $f$ is the activation function, $g$ is the locking function, $k$ is the key, $w_{i,j}^{l}$ is the weight from neuron $j$ in layer $l-1$ to neuron $i$ in layer $l$, and $b_{i}^{l}$ is the bias of neuron $i$ in layer $l$. The form of $g$ can vary, but will take on the identity function when $k=k'$, where $k'$ is the correct key value. An example of a $g(x;k)$ function for point-function locking with a fixed key value is in Fig. \ref{fig:locking_scenario}B. One or more of these locked neurons will be present in the network. 

The attack goal is to find a $k$ that minimizes loss function $\mathcal{L}$:
\begin{equation} \label{eqn:loss_func}\vspace{-2mm}
\mathcal{L}\left( \mathcal{U},\mathcal{Y},k \right) =-\sum_{j=0}^{C-1}{\mathcal{Y}_i\cdot \log \left( \mathrm{soft}\max \left( \mathcal{F}\left( \mathcal{U},k \right) _i \right) \right)},
\end{equation}
where $\mathcal{U}$ signifies the model inputs and $\mathcal{Y}$ denotes the attacker-desired output labels (i.e., output labels for trigger inputs are replaced with attacker-specified pseudo-labels corresponding to the intended adversarial function of the model (see Sec. \ref{sec:attack_method})). $\mathcal{F}$ symbolizes the complete network, and $\mathcal{F}\left( \mathcal{U},k \right) _i$ represents the output logits for class i. The $\sum{}$ component encompasses all classes under consideration. Stated formally, the goal of a hardware neural trojan attacker is to identify the adversarial trojan key, $k_{adv}$, given by:
\begin{equation} \label{eqn:opt_goal}
k_{adv} = \underset{k}{\arg\min}\:\mathcal{L}(\mathcal{U},\mathcal{Y};k)
\end{equation}

To adapt this formulation to other locked locations in the neuron, the $g$ function can be moved. For example, locking in the non-linearity, which was implemented in our attack evaluation in Sec. \ref{sec:eval}, can be represented by the equation below:
\begin{equation}
x_{i}^{l}=g\left(f\left(\sum_{j}x_{i}^{l-1}w_{i,j}^{l}+b_{i}^{l}\right);k\right)=g\left(f\left(\sum_{j}s_{i,j}^{l}+b_{i}^{l}\right);k\right)
\end{equation}
In this case, the goal of the attacker (i.e., Eqn. \ref{eqn:opt_goal}) remains the same.

% Let us assume that the probability density of $\mathbf{s}_{i}$ is given by $p(\mathbf{s}_{i})$.  Then, the expected squared difference between the neuron output with correct and incorrect keys is given as:
% \begin{equation}
% \mathbb{E}[...]=\int\limits_{\mathbb{R}^{N_{l-1}}}p(\mathbf{s}_{i}^{l})\left[f\left(\sum_{j}g\left(s_{i,j}^{l};k\right)+b_{i}^{l}\right)-f\left(\sum_{j}s_{i,j}^{l}+b_{i}^{l}\right)\right]^{2}\mathrm{d}\mathbf{s}_{i}^{l}
% \end{equation}

\vspace{-2mm} 
\subsection{Attack Methodology} \label{sec:attack_method}

To launch a hardware neural trojan attack against an arbitrary logic locked neural accelerator, the attacker must find a trojan key, $k_{adv}$, as defined by Eqn. \ref{eqn:opt_goal}. Intuitively, this is achieved by treating the locking key as a weight in the locked model and \textit{training} it through traditional back-propagation methods to optimize $k_{adv}$. We outline this approach and its implementation below:

\begin{enumerate}[leftmargin=*]
    \item A function $g(x;k)$ is formulated to represent any locked neuron in the model (see Sec. \ref{sec:attack_form}). Note that a single $k$ value can be shared among multiple locked neurons. To determine $g(x;k)$, the attacker can analyze the locked netlist to determine the output for any given input/key combination. Alternatively, the attacker can be assumed to understand the locking construction sufficiently to model $g(x;k)$ (a valid assumption under Kerckhoff's Principle). This does not leak the correct key value as it is defined by the locked netlist, which the attacker knows. 
    \item The attacker classifies each $k$ in the model as a trainable weight parameter. All traditional weights in the neural model are considered to be frozen (i.e., fixed) as no traditional weight retraining/modification can be performed by the attacker.
    \item A subclass (i.e., a trigger) is selected as the target of the attack. A set of adversarial \textit{pseudo-labels} is then defined for the network that represent the attackers intended functionality for trigger inputs in the trojan-key-compromised accelerator. The definition of these pseudo-labels differ based on attacker goal:
    \begin{enumerate}[leftmargin=14pt]
        \item \textbf{Untargeted Attack:} The adversarial pseudo-labels are altered non-directionally through random modification.
        \item \textbf{Targeted Attack:} The adversarial pseudo-labels are altered directionally, by specifying a target class for trigger inputs.
    \end{enumerate}
    The resulting set of pseudo-labels can be used to assess our target loss function (Eqn. \ref{eqn:loss_func}) for a key value.
    \item The trigger subclass and its corresponding pseudo-labels are then subjected to training where $k$, the trojan key, is the only trainable parameter in the network. Training is performed with traditional back-propagation methods. The resulting value after training, $k_{adv}$, corresponds to a trojan key that maximizes trigger input misclassification for the target locked accelerator.
\end{enumerate}

We emphasize that all model weights are frozen during the entire trojan key optimization process. Hence, our attacker does not need to know their value (i.e., white-box model access). Instead, they can run back-propagation to train the trojan key value by applying a key and observing the corresponding input/output for each neuron in the model using the scan-chain of the black-box oracle.

\vspace{-2mm} 
\section{Attack Evaluation} \label{sec:eval}

To evaluate our hardware neural trojan attack methodology, we applied it to three logic-locked neural accelerators, each implementing a different neural architecture (MLP, ResNet18 \cite{he2016deep}, and ResNeXt29 \cite{xie2017aggregated}). These architectures were trained on MNIST, CIFAR10, and CIFAR10, respectively. We locked a single hardware neuron in each accelerator, which executed one randomly selected neuron in the neural model (we relax this in Sec. \ref{sec:exp3}). In each locked neuron, we simulated point-function-style locking, such as \cite{yasin2016sarlock, xie2018anti, yasin2017provably, shakya2020cas}, that injects high error for a small set of nearby input values that are shifted based on the applied key. The locking was implemented in the non-linearity of the locked hardware neuron. To evaluate our attack, we implemented the methodology outlined in Sec. \ref{sec:attack_method} using PyTorch and performed three experiments with these benchmarks.

\vspace{-2mm} 
\subsection{Experiment 1: Untargeted Trojan Attack}\label{sec:exp1}

To evaluate our untargeted hardware neural trojan attack from Sec. \ref{sec:attack_method}, we launched it against each benchmark accelerator with two randomly selected trigger input classes (class 1 and 6). This attacker aims to identify a trojan key that causes a specified input class (i.e., the trigger) to be misclassified without degrading the classification accuracy for other classes. We have aggregated the data of this experiment in Tbl. \ref{tab:untargeted_data}. Each cell in the table highlights the change in output classification accuracy for the test set caused by the trojan key (i.e., the trojan key accelerator accuracy minus the unlocked accelerator accuracy for each class). In each case, a trojan key was identified that greatly reduced classification accuracy for the trigger input class, while not substantially impacting the accuracy of other classes. For the largest model (ResNeXt29), our attack found a trojan key that reduced the trigger input class accuracy by 74\% with only a 1.7\% accuracy degradation in other classes on average.

\begin{table}[t]
     \centering
     \small
     \renewcommand{\arraystretch}{0.9} % Default value: 1

     \begin{tabular}{c c|c|c|c|c|c|c}
         \multicolumn{2}{r}{} & \multicolumn{2}{c}{\textbf{MLP}} & \multicolumn{2}{c}{\textbf{ResNet18}} & \multicolumn{2}{c}{\textbf{ResNeXt29}} \\ 
         \cmidrule(lr){3-4}\cmidrule(lr){5-6}\cmidrule(lr){7-8}
         \multicolumn{2}{r}{\textbf{Trigger Class:}} & \multicolumn{1}{c}{1} & \multicolumn{1}{c}{6} & \multicolumn{1}{c}{1} & \multicolumn{1}{c}{6} & \multicolumn{1}{c}{1} & \multicolumn{1}{c}{6} \\
         \hline
         \parbox[t]{2mm}{\multirow{10}{0pt}{\rotatebox[origin=c]{90}{\textbf{Output Class}}}}&\textbf{Class 0} & 0.1 & 0.3 & -5.2 & 1.7 & -4.5 & 0.9 \\
         \cline{2-8}
         &\textbf{Class 1} & \cellcolor{red!25} -37.4 & -0.6 & \cellcolor{red!25} -50.6 & 0.4 & \cellcolor{red!25} -69.5 & -0.6 \\
         \cline{2-8}
         &\textbf{Class 2} & -3.1 & -3.9 & -0.1 & -3.1 & -1.7 & -4  \\
         \cline{2-8}
         &\textbf{Class 3} & -3.6 & -1.7 & -1.2 & -2.2 & 0.5 & -5.9 \\
         \cline{2-8}
         &\textbf{Class 4} & 0.2 & -0.6 & -1.6 & -4.9 & -0.4 & -5.3 \\
         \cline{2-8}
         &\textbf{Class 5} & 0.6 & 0.3 & 0.9 & -1.4 & -0.5 & 1 \\
         \cline{2-8}
         &\textbf{Class 6} & 0 & \cellcolor{red!25} -41.9 & -0.9 & \cellcolor{red!25} -65 & 0.4 & \cellcolor{red!25} -78.6 \\
         \cline{2-8}
         &\textbf{Class 7} & -0.8 & 0 & -2.5 & -0.8 & -2 & -0.4  \\
         \cline{2-8}
         &\textbf{Class 8} & 0.6 & -7.4 & -3.9 & -2.3 & -1.8 & -1.5 \\
         \cline{2-8}
         &\textbf{Class 9} & -22 & 14.3 & -7.5 & -1.2 & -5 & 0.3 \\
         \hline
        
     \end{tabular}

     \vspace{1mm}
     \caption{Impact of untargeted hardware neural trojan attack on benchmark logic-locked accelerators. Each column represents an attack scenario (i.e., an architecture and trigger input class combination). Each row represents an output class. Column/row intersections contain the difference in accelerator accuracy caused by a trojan key for each class (determined by the row) and attack scenario (determined by the column) compared to an unlocked accelerator. The cell corresponding to the trigger input for each attack scenario is colored red.\vspace{-1mm}}
     \label{tab:untargeted_data}
\end{table}

\vspace{-2mm} 
\subsection{Experiment 2: Targeted Attack}

To evaluate our targeted hardware neural trojan attack, we launched it against the ResNet18 accelerator with two randomly selected target output classes (class 0 and 9) for the two trigger input classes from Sec. \ref{sec:exp1} (class 1 and 6). This attacker aims to identify a trojan key that causes a specified input class (i.e., the trigger) to be misclassified to a specific output class (i.e., the target) without degrading the classification accuracy for other classes. We have aggregated the data of this experiment in Tbl. \ref{tab:targeted_data}. Each cell in the table contains the probability that an input from the trigger input class will be classified to each possible output class. While a trojan key was identified that did improve the likelihood of a trigger input being classified to the target class, the resulting increase was quite small (only $1.15\%$ on average). This indicates limitations in the targetability of our proposed attack methodology. This makes sense given that the attacker can control when the error is injected, but has no control over the magnitude of the injected error. This prohibits the attacker from tuning the error to produce targeted misclassification.

\begin{table}[t]
     \centering
     \small
     \renewcommand{\arraystretch}{0.9} % Default value: 1

     \begin{tabular}{c c|c|c|c|c|c|c}
         \multicolumn{2}{c}{} & \multicolumn{3}{c}{\textbf{Trigger: Class 1}} & \multicolumn{3}{c}{\textbf{Trigger: Class 6}} \\ 
         \cmidrule(lr){3-5}\cmidrule(lr){6-8}
         \multicolumn{2}{r}{\textbf{Target Class:}} & \multicolumn{1}{c}{None} & \multicolumn{1}{c}{0} & \multicolumn{1}{c}{9} & \multicolumn{1}{c}{None} & \multicolumn{1}{c}{0} & \multicolumn{1}{c}{9} \\
         \cline{1-8}
         \parbox[t]{2mm}{\multirow{10}{0pt}{\rotatebox[origin=c]{90}{\textbf{Output Class}}}}&\textbf{0} & 4.4 & \cellcolor{green!25} 4.5 & 4.3 & 6 & \cellcolor{green!25} 5.9 & 5.6 \\
         \cline{2-8}
         &\textbf{1} & \cellcolor{red!25} 47.3 & \cellcolor{red!25} 45.1 & \cellcolor{red!25} 44.3 & 6.2 & 5.1 & 5.5 \\
         \cline{2-8}
         &\textbf{2} & 1.2 & 1.4 & 1.8 & 11.3 & 10.4 & 9.6  \\
         \cline{2-8}
         &\textbf{3} & 6.4 & 8 & 5.3 & 13.1 & 13.8 & 12.1 \\
         \cline{2-8}
         &\textbf{4} & 2.5 & 2.8 & 2.5 & 8.5 & 6.6 & 7.7 \\
         \cline{2-8}
         &\textbf{5} & 3.5 & 3.5 & 3.5 & 8.4 & 8.1 & 7.1 \\
         \cline{2-8}
         &\textbf{6} & 9.6 & 8.6 & 9.9 & \cellcolor{red!25} 32 & \cellcolor{red!25} 33 & \cellcolor{red!25} 34.4 \\
         \cline{2-8}
         &\textbf{7} & 3.5 & 3.8 & 3 & 3.1 & 3 & 2.9  \\
         \cline{2-8}
         &\textbf{8} & 9 & 9.6 & 10.9 & 5.8 & 6.8 & 6.8 \\
         \cline{2-8}
         &\textbf{9} & 12.6 & 12.7 & \cellcolor{green!25} 14.5 & 5.6 & 7.3 & \cellcolor{green!25} 8.3 \\
         \hline
        
     \end{tabular}
     \vspace{1mm}
     \caption{Impact of targeted hardware neural trojan attack on the locked ResNet18 accelerator. Each column represents one attack scenario (i.e., a trigger input and target output class combination). Each row represents an output class. A column/row intersection contains the probability that the trojan key from a specific attack scenario (determined by the column) causes the accelerator to classify a trigger input from the test set to each possible output class (determined by the row). The cell corresponding to the trigger input for each attack scenario is red; the target output class is green.\vspace{-1mm}}
     \label{tab:targeted_data}

\end{table}

\vspace{-2mm} 
\subsection{Experiment 3: Multiple Locked Neurons} \label{sec:exp3}

\begin{figure}[b]
\centerline{\includegraphics[trim={0.2cm 0.4cm 0.6cm 4.4cm},clip,scale=.55]{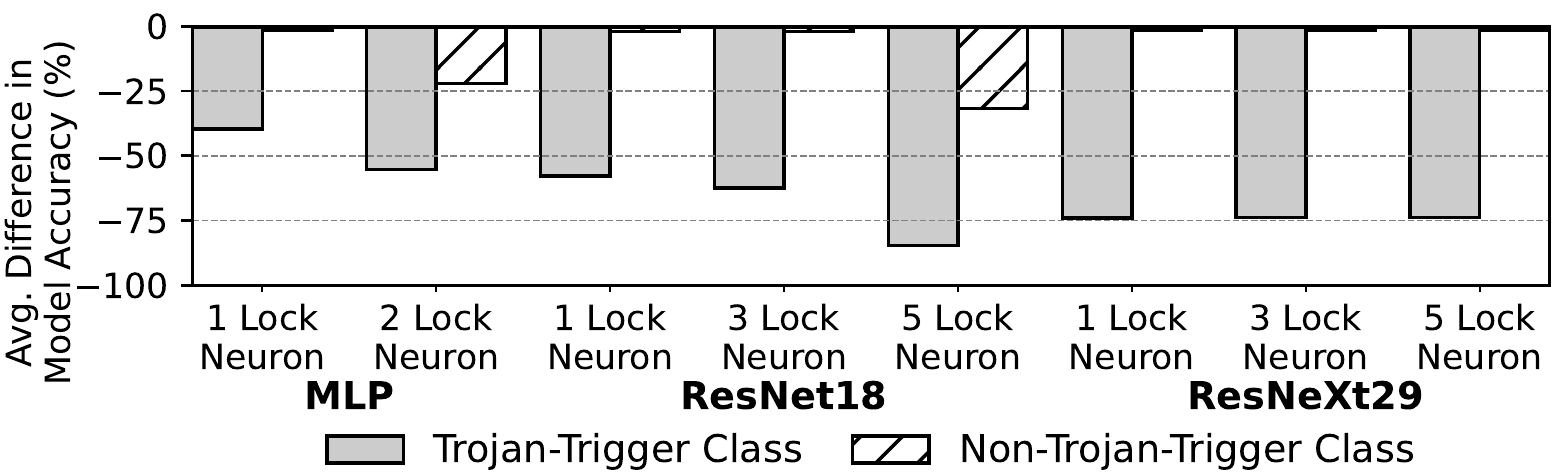}}
\caption{Impact of the number of model neurons mapped to a locked hardware neuron on trojan key effectiveness.}
\label{fig:neuron_scale}
\end{figure}

Finally, we consider the case where more than one model neuron is mapped to the same locked hardware neuron, allowing the attacker to affect multiple model neurons with a single key. This mimics a scenario where a single locked hardware resource is used to execute multiple neurons in the model. To evaluate this scenario, we ran our untargeted attack methodology on our three accelerator benchmarks with a different number of randomly selected model neurons mapped to the locked hardware neuron. For our CNN architectures, we evaluated 1, 3, and 5 model neurons mapped to the locked neuron. For the much smaller MLP architecture, we evaluated only the 1 and 2 model neuron case. We aggregated the results of this experiment for our two trigger input classes from Sec. \ref{sec:exp1}. The resulting degradation in model accuracy caused by the trojan key compared to an unlocked accelerator is in Fig. \ref{fig:neuron_scale}. For smaller models (e.g., the MLP), the likelihood of non-trojan-trigger inputs being misclassified increased as the number of model neurons mapped to the locked hardware neuron increased. However, for our largest model, ResNeXt29, trojan key performance remained constant. Regardless of the number of model neurons mapped to the locked hardware neuron in ResNeXt29, trigger input classification accuracy dropped by $74\%$ with only a $1.7\%$ reduction in the classification accuracy of other input classes. This indicates that our attack remains effective against larger models (e.g., ResNeXt29), which are more likely to use a single hardware resource to execute multiple model neurons than a smaller model (e.g., the MLP).

\vspace{-2mm} 
\subsection{Discussion and Future Work}

Our proposed attack methodology was able to successfully identify trojan keys in a variety of neural accelerator configurations, neural architectures, and neural models, demonstrating the feasibility of our attacker. However, this evaluation is by no means exhaustive given the huge variety of logic locking techniques, neural accelerators, neural architectures, and neural models proposed. We highlight three areas that were not explored in our experimental evaluation as promising directions for future work.

\begin{enumerate}[leftmargin=*]
    \item \textbf{Stealthy trojan triggers.} We considered an attacker that caused one input class (trigger) to misclassify as another (target), while otherwise limiting model accuracy degradation. While this shows the feasibility of our attacker, the trigger of the resulting hardware neural trojan has limited stealthiness. A more potent attack could target input features present only in a small subset of inputs, producing trojan keys with stealthier triggers. 
    \item \textbf{Alternative logic locking families.} While many state-of-the-art logic locking schemes employ a point-function-based approach, there are other prominent families that distribute error throughout the input space differently. Different error distributions can impact the effectiveness of hardware neural trojans or enable new trojan capabilities.
    \item \textbf{Deep neural network (DNN) architectures.} The evaluated neural architectures are shallow compared to DNNs that dominate the state-of-the-art. The size of DNN models would have two consequences. 1) The impact of any one neuron on classification accuracy is likely limited. 2) In a deep learning accelerator, a large number of neurons are likely to be mapped to each hardware neuron. Therefore, while the trojan key search problem may become more complex in deeper models, the possible expressiveness of trojan keys may also increase to allow more potent hardware neural trojans to be identified.
\end{enumerate}

\vspace{-2mm} 
\section{Conclusion}

We proposed a novel untrusted foundry attacker on logic-locked neural accelerators. This attacker inserts a neural trojan into a neural accelerator by exploiting the corruption caused by logic locking when a wrong key is applied. This results in a compromised neural accelerator that can be sold on the gray market. Given the safety-critical nature of many neural accelerator applications (e.g., autonomous driving), such an attack could have severe consequences. We developed a theoretically-robust methodology to launch this attack on arbitrary logic-locked neural accelerators and evaluated it in several benchmarks. In each benchmark accelerator, our attack successfully identified \textit{trojan keys} capable of producing misclassifications for an attacker-specified trigger input class with minimal impact on network accuracy for other non-trigger inputs.

\vspace{-2mm} 
\section{Acknowledgements}
This material is based on research sponsored by the National Science Foundation under grant number 2245573 and the Air Force Research Laboratory under agreement number FA8750-20-2-0503. The U.S. Government is authorized to reproduce and distribute reprints for Governmental purposes notwithstanding any copyright notation hereon. The views and conclusions contained herein are those of the authors and should not be interpreted as necessarily representing the official policies or endorsements, either expressed or implied, of the Air Force Research Laboratory or the U.S. Government.

%%
%% The next two lines define the bibliography style to be used, and
%% the bibliography file.
% then add the following to balance the last page (2 even length columns).
% If you have used \usepackage{balance} include \balance between \bibliographystyle & \bibliography
\bibliographystyle{ACM-Reference-Format}
\balance
\bibliography{sample-base, ref}

%%
%% If your work has an appendix, this is the place to put it.

\end{document}